\documentclass{aa}
\usepackage{graphicx}
\usepackage{txfonts}
\usepackage{natbib}
\bibpunct{(}{)}{;}{a}{}{,}

\begin{document}

\title{Modal Filtering for Nulling Interferometry}

\subtitle{First Single-Mode Conductive Waveguides in the Mid-Infrared}

\author{L. Labadie,\inst{1}
        P. Labeye,\inst{2}
        P. Kern,\inst{1}
        I. Schanen\inst{3}
        B. Arezki\inst{1}
        J.-E. Broquin\inst{3}}

\offprints{L. Labadie\\
           \email{Lucas.Labadie@obs.ujf-grenoble.fr}}

\institute{Laboratoire d'Astrophysique de l'Observatoire de Grenoble, BP53, 38041 Grenoble Cedex 9, France
           \and
           Laboratoire d'Electronique et des Technologies de l'Information (LETI),CEA-DRT-LETI,17 rue des Martyrs, 38054 Grenoble C\'edex 9, France
           \and
           Institut de MicroElectronique et Photonique, UMR 5130, 23 rue des Martyrs, BP257,38016  Grenoble C\'edex 1, France
           }

\date{Received September 15, 1996; accepted March 16, 1997}

\abstract{This paper presents the work achieved for the manufacturing and characterization of first single-mode waveguides to be used as modal filters for nulling interferometry in the mid-infrared range [4-20 $\mu$m]. As very high dynamic range is mandatory for detection of Earth-like planets, modal filtering is one of the most stringent instrumental aspects.
The hollow metallic waveguides (HMW) presented here are manufactured using micro-machining techniques.
Single-mode behavior has been investigated in laboratory through a technique of polarization analysis while transmission features have been measured using flux relative comparison.
The single-mode behavior have been assessed at $\lambda$=10.6 $\mu$m for rectangular waveguides with dimensions $a$=10 $\mu$m and $b$$\le$5.3 $\mu$m with an accuracy of $\sim$2.5\%. The tests have shown that a single-polarization state can be maintained in the waveguide. A comparison with results on multi-mode HMW is proposed. Excess losses of 2.4 dB ($\sim$ 58\% transmission) have been measured for a single-mode waveguide. In particular, the importance of coupling conditions into the waveguide is emphasized here.
The goal of manufacturing and characterizing the first single-mode HMW for the mid-infrared has been achieved. This opens the road to the use of integrated optics for interferometry in the mentioned spectral range.
}
\authorrunning{L.Labadie et al.}
\titlerunning{First Single-Mode Waveguides for the Mid-Infrared}
\maketitle

\section{Introduction}\label{intro}

Since the discovery of a Jupiter-mass companion around 51 Peg by \citet{Mayor}, the next challenges that have been addressed by modern astronomy are the search for habitable telluric planets and the search of indices of life on them. The main challenge being the huge ratio of star to planet flux, \citet{Bracewell} and \citet{Angel86} have suggested to observe in the mid-infrared range. Indeed, in the case of a Sun/Earth system the luminosity ratio is about 10$^{9}$ in the visible while it drops to 10$^{6}$ at 10 $\mu$m. The latter authors have shown that reliable biomarkers such as H$_{2}$O ($<$ 8 $\mu$m), O$_{3}$ (9.6 $\mu$m) and CO$_{2}$ (15 $\mu$m) could be researched by spectroscopy in the mid-infrared.
At those wavelength, the search for Earth-like planets could use an infrared nulling interferometer in space operating in the [4 - 20 $\mu$m] band \citep{Bracewell2}. A nulling interferometer generates destructive interferences for an on-axis star by introducing an achromatic $\pi$ phase shift between the arms of the interferometer, which results in the complete rejection of the starlight. Aside, for a well-tuned configuration of the baseline, the signal from an off-axis planet will be transmitted since it interferes constructively at the output of the instrument. Thus, this type of instrument provides at the same time high angular resolution through baselines of several tens of meters and high dynamic range through the intrinsic coronographic feature of its transmission map \citep{Leger96}.\\
However, reaching a rejection ratio of 10$^{6}$ is subject to three major limitations which are: phase defects that includes residual OPD errors, pointing errors, optics defects and micro-roughness that induce high-frequency defects; Overall and local amplitude shifts; Polarization mismatching due to the rotation of the electric field in the optical train.
Concerning the first two points (excepted for residual OPD), authors have shown that leaks could be maintained below 10$^{-6}$ by implementing spatial filtering using either pinholes \citep{Ollivier} or single-mode waveguides \citet{Mennesson}.
Polarization mismatches could be corrected using polarization maintaining devices.\\
Filtering with single-mode waveguides is already largely used in near-infrared interferometry. This is done with optical fibers \citep{Foresto92} or single-mode integrated optics (IO) \citep{Kern}. The later option provides, in addition to modal filtering, optical functions like beam combination which is integrated on a small size optical chip. This solution relaxes instrumental constraints like alignment, mechanical and thermal stability \citep{Malbet99} and provides to date calibrated astrophysical data from several interferometric facilities \citep{Kervella}.\\
The importance of modal filtering is now commonly accepted. In this context of mid-infrared nulling interferometry, the lack of single-mode waveguides has triggered specific studies to develop mid-infrared chalcogenide fibers \citep{Borde03} and silver-halide fibers \citep{Wallner}. \citet{Wehmeier} has also started theoretical studies on conductive waveguides for the mid-infrared.\\
The present work aims to conduct the extension of single-mode integrated optics concepts to the mid-infrared range, where it may be advantageous for future space-based interferometers like Darwin \citep{Fridlund} or TPF \citep{Beichman}.
In this paper, we present the promising results on the first manufactured single-mode Hollow Metallic Waveguides (HMW) for the mid-infrared range. HMW is a totally new approach in the implementation of guided optics properties for mid-infrared interferometry with respect to dielectric technologies. Although theory can help us to predict the performances of a new component, the novelty of the technology process brings several unknowns as surface quality, coupling and propagation losses on the final performances. Only first fabrication and validation tests can provide reliable conclusions that can help us to improve the performances by successive iterations.\\
First, we recall quantitative results on the effects of using single-mode waveguides for filtering.
Secondly, we describe the impact of HWM geometry on the single-mode behaviour as well as the applied manufacturing process. Then, we present the laboratory workbench built for the characterization phase and the adopted procedure for the specific measurements. In a fourth section are presented the results on the modal behavior and the transmission features of the manufactured HMW. Finally, we discuss the results and the hollow metallic waveguide properties.

\section{Modal Filtering with Single-Mode Waveguides}\label{filtering}

\begin{figure*}
\begin{minipage}{\textwidth}
\centering
\includegraphics[width=6.5cm]{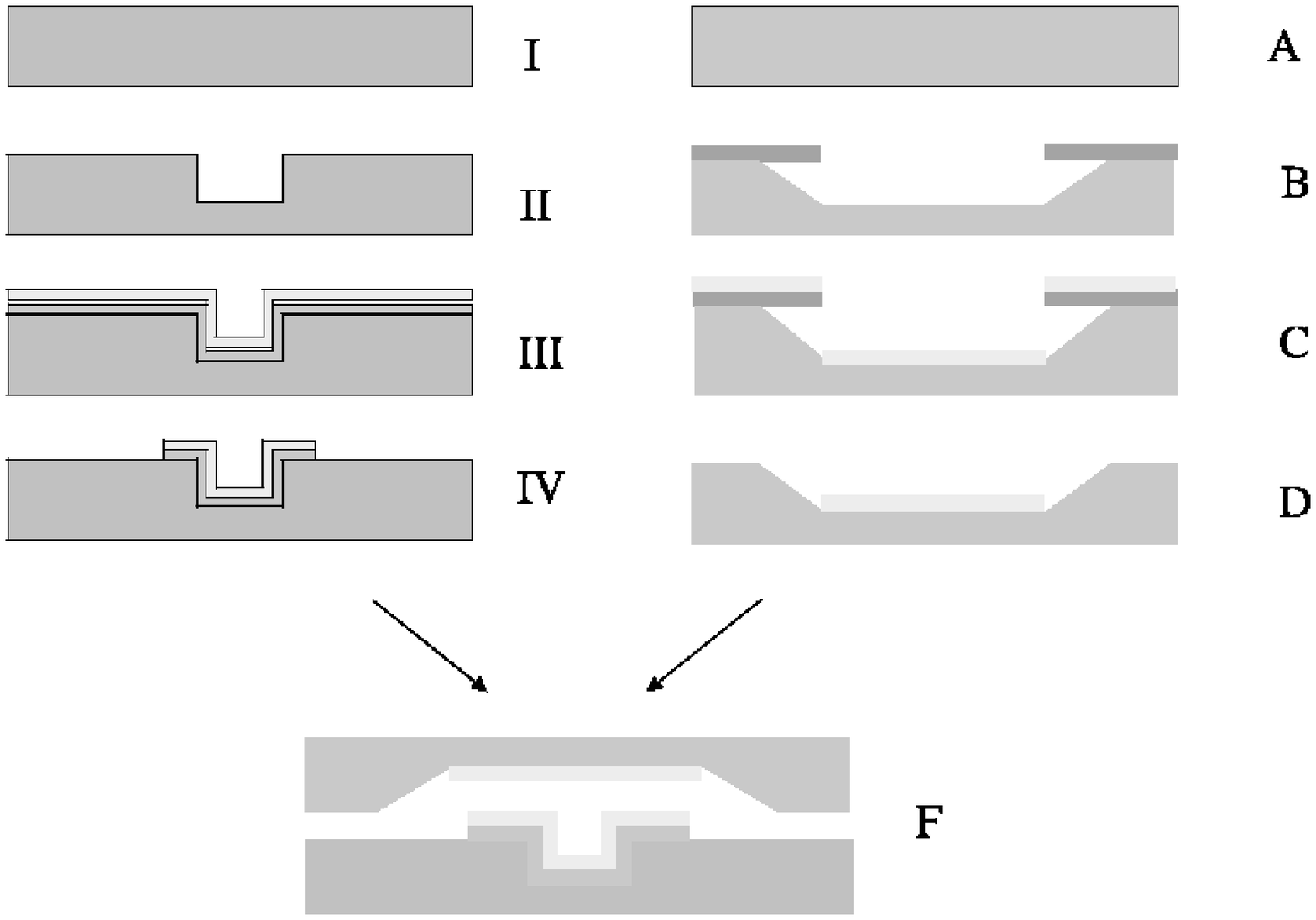}
\hspace{1cm}
\includegraphics[width=6.0cm]{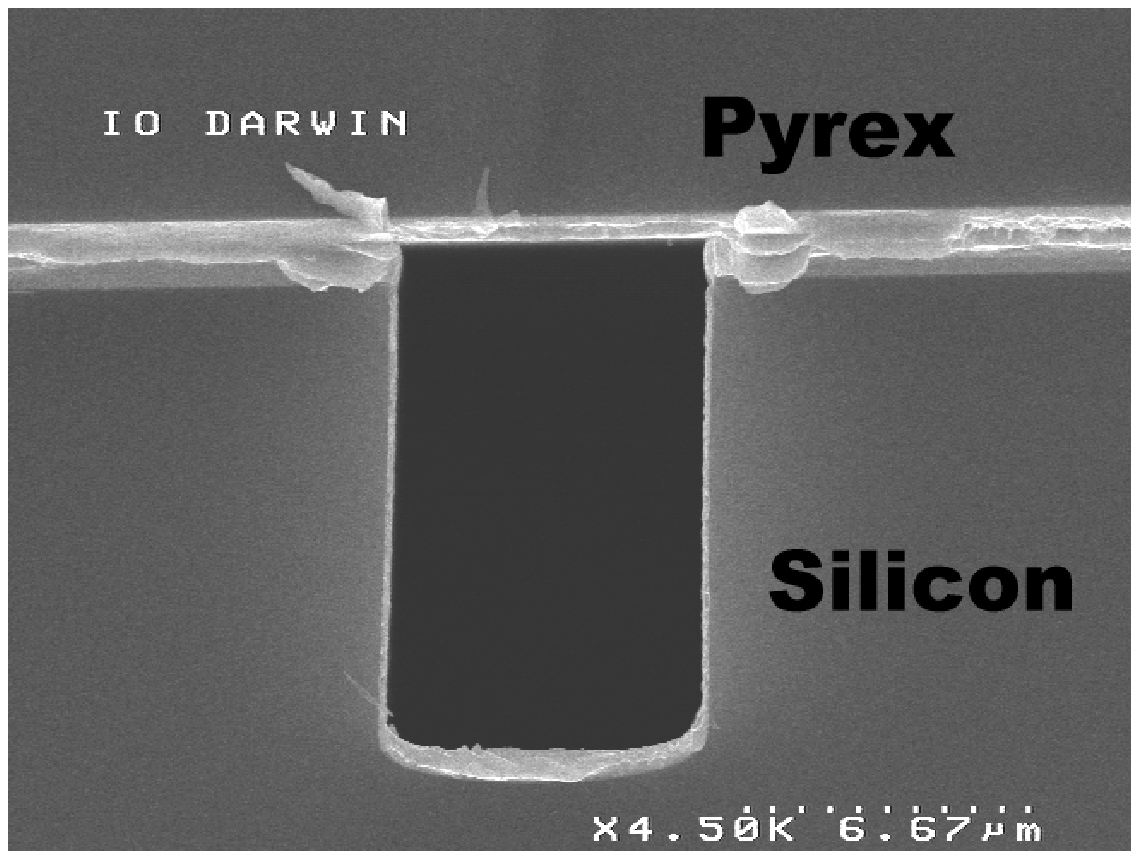}
\end{minipage}
\\
\caption{Left panel: Process flow of HMW manufacturing.I: Silicon 4-inch wafer - II: Guide definition by photo-lithography and RIE - III: Thermal oxidation and gold deposition - IV: Gold and silica wet etching - A: $Pyrex^{TM}$ 4-inch wafer - B: Photo-lithography and $Pyrex^{TM}$ wet etching - C: Gold deposition - D: Photo-resist stripping - F: Anodic bonding. Right panel: photograph of a HMW input using Scanning Electron Microscope. The Pyrex cover is maintained by anodic bonding on the silicon substrate. The etching depth is equal to $10\mu m$. The gold deposition is thicker on the bottom part of the waveguide with respect to the lateral walls.}\label{building_process}
\end{figure*}

In order to achieve a 10$^{5}$ rejection ratio with a four-telescope configuration, \citet{Leger95} pointed out the following constraints on the instrument:
\begin{itemize}
\item Defects due to optical quality and micro-roughness as well as residual OPD should be kept smaller than $\lambda$/2000 at 10 $\mu$m (i.e. $\lambda$/100 in the visible at 0.5 $\mu$m).
\item Pointing errors should remain below 1/80 of the Airy disk in the visible, or 1/1600 at 10 $\mu$m.
\end{itemize}
\noindent To relax those constraints, \citet{Ollivier} has first proposed to use pinholes to be placed at the focus of each telescopes. A pinhole with the size of the Airy disk filters out the small-scale defects (i.e. the high spatial frequencies) providing efficient spatial-cleaning. Defects due to dust and polishing are well corrected with a rejection increase of three orders of magnitude. For instance, pinhole filtering permits to reach a rejection rate of 10$^{6}$ when the transmission is degraded by 1\% due to dust, while it remains below 10$^{3}$ without filtering.
Pinholes efficiency, however, is sensitively reduced for large-scale defects. In particular tilt errors are not significantly corrected with this method and the rejection ratio is improved by less than a factor two. Finally, residual OPD errors cannot be corrected by any type of spatial filtering since they do not affect the intensity distribution of the Airy pattern in the Fourier plane at the telescope focus.\\
\noindent Further improvements are achieved using single-mode waveguides to implement modal filtering. In guided optics theory \citep{Marcuse}, a single-mode waveguide does propagate only one mode, the fundamental one, whose amplitude spatial distribution is only constrained by the geometrical and optical parameters of the waveguide, independently of the incoming wavefront. Coupling a corrugated wavefront on the fundamental mode converts the input phase fluctuations into overall intensity fluctuations of the propagating mode. The resulting wavefront at the waveguide output is then cleaned from all phase defects. When placed at the focus of each telescope, the fringe visibility is then only affected by amplitude mismatching which is less drastic than phase defects. However, single-mode waveguides provide neither a correction of the residual OPD nor a correction of overall amplitude unbalancing. Those errors can be corrected with classical fringe tracking and active amplitude matching systems.\\
\citet{Mennesson} provided a quantitative comparison of the optical constraints set respectively by no filtering mode, pinhole filtering and single-mode waveguide filtering to reach a typical rejection ratio of 10$^6$. Concerning high-order optics defects and micro-roughness, the wavefront quality at $\lambda$=10 $\mu$m drops from $\lambda$/4400 rms without spatial filtering to $\lambda$/400 rms with pinhole filters and to $\lambda$/63 rms with a single-mode waveguide, that is a gain of almost two order of magnitude.
For the telescopes pointing errors, single-mode waveguides become once again very efficient. Reaching a deep null of 10$^6$ at 10 $\mu$m requires a control of the tip-tilt up to 1.2 mas rms for 1.5m-class telescopes. In the same conditions, modal filtering with single-mode waveguides relaxes this constraint to 38 mas rms.
Local amplitude errors due, for instance, to imperfect coatings reflectivity can also be corrected more efficiently with single-mode waveguides. For the same previous rejection ratio, the relative local amplitude shift to be monitored goes from 0.2\% without filtering to 2\% with pinhole filtering and to 10\% with single-mode filtering.\\
Concerning polarization mismatching, a way to get rid of this effect is to implement single-mode and single-polarization waveguides in the optical train. This aspect is discussed in Sect.~\ref{results}.\\
Finally, modal filtering can be implemented after coherent recombination of the beams using only one waveguide rather than placing a waveguide at each telescope focus. The advantage is that any additional phase corruption after filtering is prevented as well as any differential effect due to slightly different waveguides.\\
\noindent Modal filtering based on the implementation of single-mode waveguides presents an evident advantage for nulling interferometry where high rejection rates are strongly constrained by the presented factors. Single-mode waveguides are preferred to pinholes since they are efficient for high-order and low-order defects, while the later devices are efficient only for small-scale defects. Therefore, the developing and manufacturing of single-mode waveguides working in the full band [4 - 20 $\mu$m] will provide fundamental improvements for nulling interferometry missions like Darwin and TPF.

\section{Geometry and Manufacturing of the Waveguides}\label{component}

A proposed solution to develop mid-infrared modal filters are hollow
metallic waveguides (HMW), which benefit of a strong heritage from
microwave engineering and which are widely used in the radioastronomy
field. Light propagation with HMW is based on multiple metallic
reflections of the radiation inside the cavity. HMW have been selected
as a candidate because they are able to cover the full Darwin band [4
- 20 $\mu$m] although this is done through the implementation of three
or four single-mode sub-bands. The question of the full coverage of
the band is similarly addressed when using dielectric solutions since
this reduces the number of infrared glassy materials that can be
exploited.\\
Within the frame of our study, the adopted geometry for our devices is
a rectangular shape (see Fig.~\ref{Waveguide_Geom}) with sides $a$ and
$b$ (and $a \ge b$).
Based our study on well established microwave models \citep{Rizzi}, we
have computed the electric field distributions of the different
propagated mode as well as their respective cut-off wavelength
(i.e. the wavelength above which the mode is not propagated
anymore). Thus, for a waveguide geometry fulfilling the condition
$a$=2$b$, the fundamental mode is the linearly polarized TE$_{\rm 10}$
with a cut-off wavelength $\lambda_{\rm c ({\rm TE_{\rm 10}})}$ =
2a. The first higher-order mode is the mode TE$_{\rm 01}$ with a
cut-off wavelength $\lambda_{\rm c({\rm TE_{\rm 01}})}$ = a and with a
linear polarization orthogonal to the TE$_{\rm 10}$ fundamental
mode. Consequently, a waveguide with the specified geometry has a
single-mode range defined by those two cut-off wavelengths, that is
$a$ $< \lambda <$ 2$a$. Note that the single-mode range remains
unchanged with a geometry verifying $b \le \frac{a}{2}$ while it
becomes 2$b$ $< \lambda <$ 2$a$ when $b \ge \frac{a}{2}$. It is then
possible to tune the single-mode range of the rectangular waveguide by
changing its geometry.
Under those conditions, we have fixed for HMW an etching depth of $a$
= 10 $\mu$m while their width ranges from $b\approx$ 4.5 $\mu$m to
$b\approx$ 10 $\mu$m (see Table~\ref{ztable}). Considering what
previously said, this geometry theoretically ensures a single-mode
behavior at $\lambda$=10.6 $\mu$m for those waveguides that comply
with the condition $b\le$5.3$\mu$m.
The waveguides fulfilling the opposite condition $b\ge$5.3$\mu$m will
have a multi-mode behavior at 10.6 $\mu$m. Our work, based on
characterization experiments at 10.6 $\mu$m, aims then to enhance
experimentally the difference in behavior between single-mode type and
multi-mode type waveguides.
\\
The manufacturing of a hollow metallic waveguide is based on a
standard micro-technology etching process of silicon substrate plus
anodic bonding of a Pyrex cover on the silicon structure. The process
is described in Fig.~\ref{building_process}.
The result is a chip containing rectangular waveguides 
spaced by 300 $\mu$m with sub-micron accuracy, with inside walls
coated with gold. The total length of a waveguide is 1 mm.
To avoid a direct transmission through the silicon substrate, the
external facets of the chip are also coated with gold by evaporation
process.\\
\noindent The results reported in this paper are related to channel waveguides both with and without tapers. A taper is a smooth transition between a multi-mode channel waveguide and a single-mode one, similar to impedance cornets in millimeter astronomy. This structure added at the input and output of the waveguide relaxes the coupling constraints. The dimensions of the taper are 40 $\mu$m width by 10 $\mu$m depth (see Fig.~\ref{Waveguide_Geom}). The 1-mm length waveguides includes the tapers when they are added.\\
\noindent The right panel of Fig.~\ref{building_process} presents a Scanning Electron Microscope (SEM) image of a 5 $\mu$m $\times$ 10 $\mu$m component. Table~\ref{ztable} summarizes the geometrical parameters of the tested components and provides the reference names used through the paper. All the waveguides are described with the reference Gxx where xx is a specific number. The letter "T" is added if the waveguide supports a taper.

\section{Laboratory Validation}\label{setup}

\begin{figure}
\centering
\includegraphics[width=8.5cm]{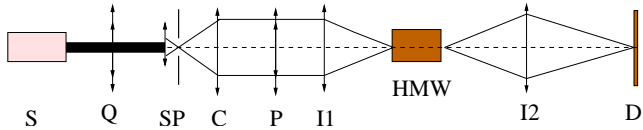}
\caption{Layout of the injection workbench at $\lambda$=10.6 $\mu m$. S: $CO_{2}$ laser source - Q: Quarter wavelength ZnSe plate - SP: 50 $\mu m$ pinhole spatial filter - P: Grid polarizer - $I_{1}$: Injection lens - HMW: Hollow Metallic Waveguide - $I_{2}$ Imaging lens - D: Detector array. The polarizer P can be placed after the component.}\label{layout}
\end{figure}

\begin{figure*}
\begin{minipage}{\textwidth}
\centering
\includegraphics[width=5.2cm]{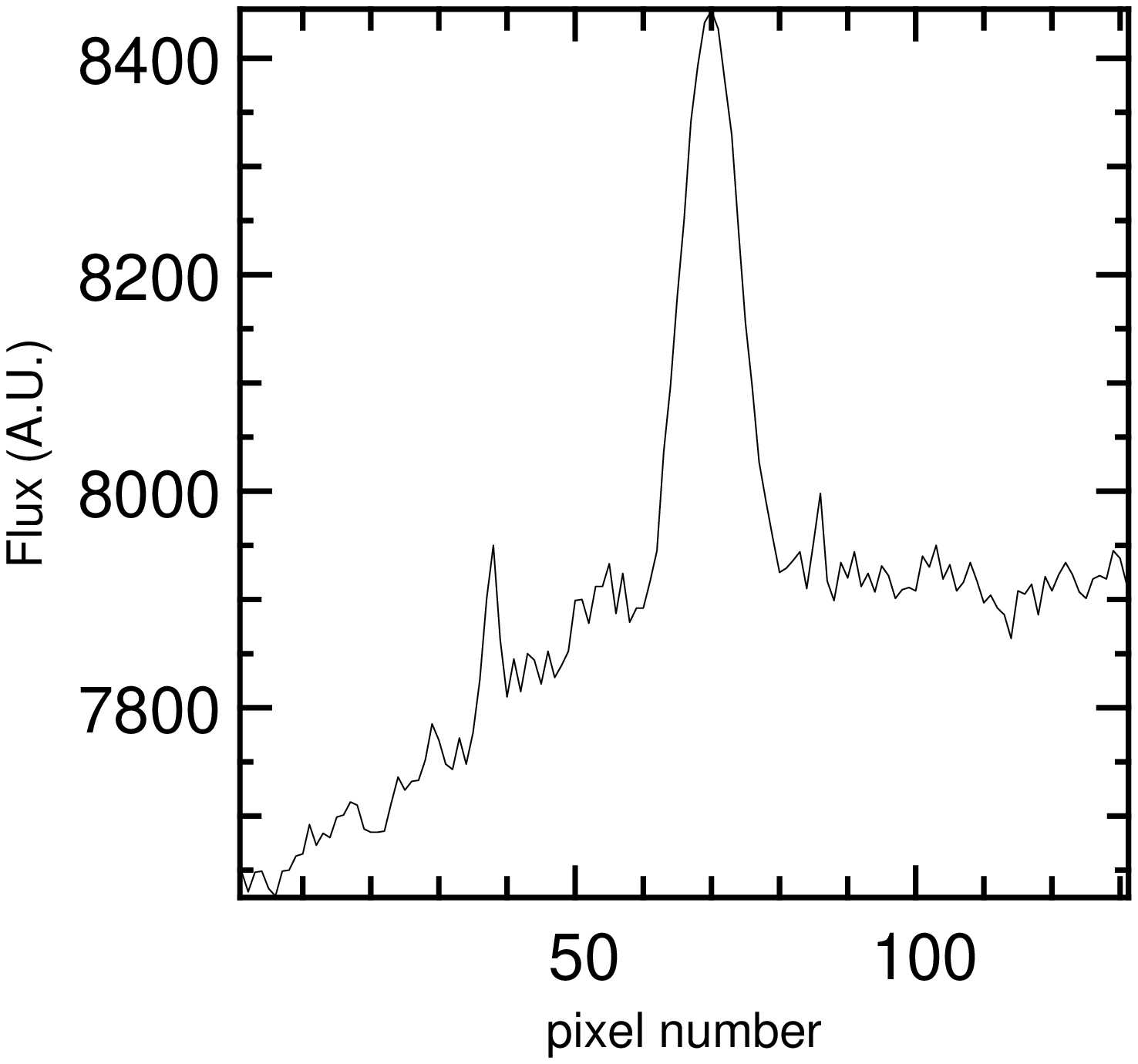}
\hspace{0.5cm}
\includegraphics[width=5.0cm]{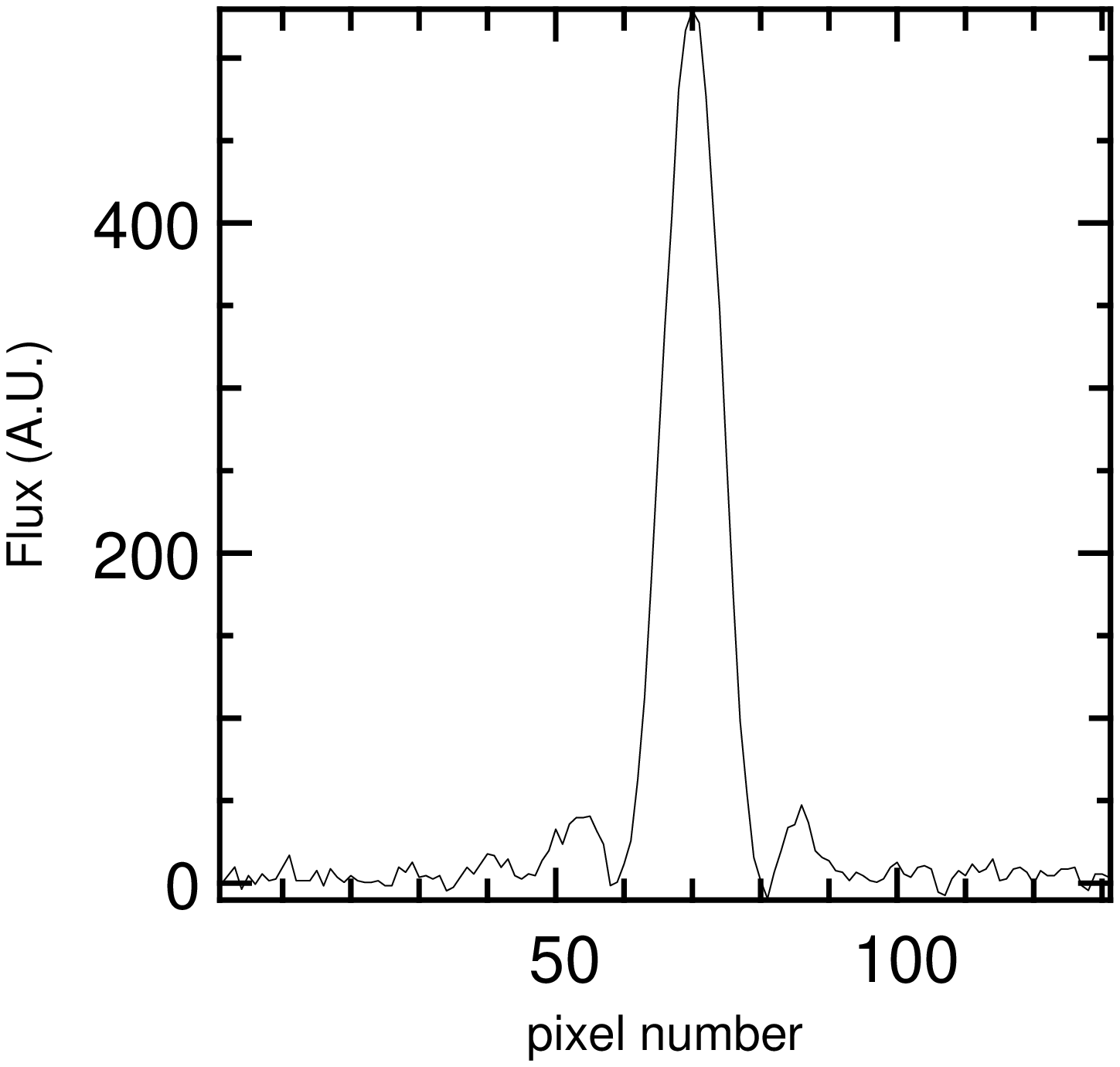}
\hspace{0.5cm}
\includegraphics[width=5.0cm]{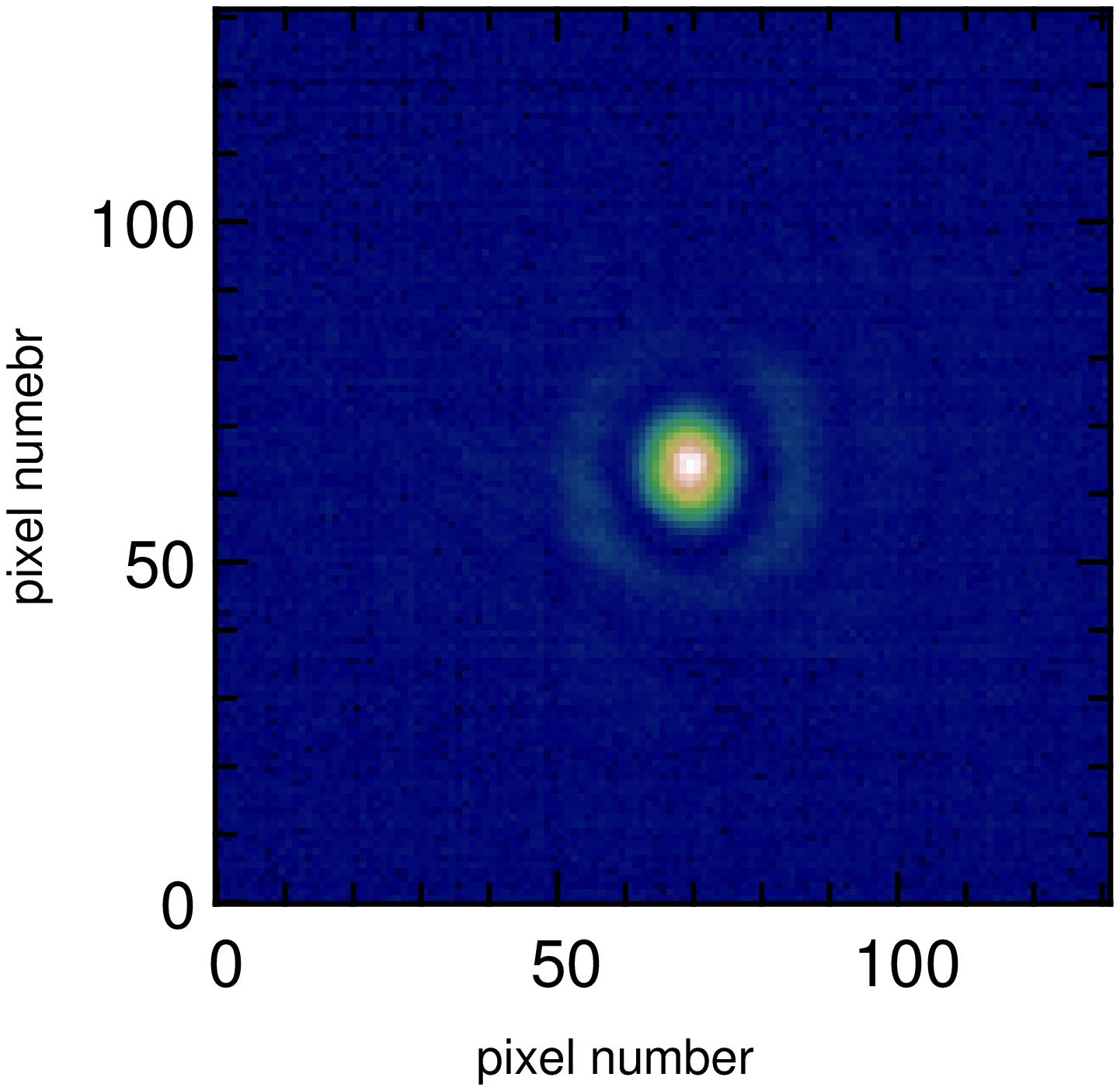}
\end{minipage}
\caption{Signal correction with flat and background suppression. The left panel of the figure shows the raw signal from the camera, which contains artifacts due to the response of each pixel. In the center panel, the correction described in Sect.~\ref{setup} has been applied, which results in a uniform distribution around the central peak. Once the background mean level has been subtracted, the signal is affected by its zero-mean Poisson noise. The right panel of the figure presents the diffraction-limited spot corresponding to the output flux of the waveguide.}\label{cor}
\end{figure*}

The laboratory workbench aims at testing and analyzing the single-mode behavior and the propagation features of the manufactured waveguides. The optical layout of the injection workbench is provided in Fig.~\ref{layout}. The chip containing the channel waveguides is supported on a three-axis positionner with fine step resolution ($\sim$1 $\mu$m) to co-align the component with the optical axis. Considering the dimensions of the waveguide aperture, a major constraint is put on the optical quality of the input beam. Therefore, we use an aspherical fast optics with numerical aperture f/1.15 to focalize the beam on the waveguide aperture, providing a spot with 30 $\mu$m full-width. A beam-expander permits to fill the 22-mm clear aperture of the focalization lens. The output flux is collected with a f/1.5 lens that images the waveguide output on the camera focal plane. The detection stage supports an uncooled 320$\times$240 micro-bolometer array detector sensitive from 8 $\mu$m to 12 $\mu$m. A grid polarizer is placed either before the component (i.e. into the collimated beam) or after the component to select the desired polarization to be analyzed. The extinction ratio of the polarizer at $\lambda$=10 $\mu$m is 185:1, which is equivalent to a polarization degree of 99.4\%. The Gaussian beam distribution of the laser is highly flattened since the collimator C with $\it{f}$=254 mm is overfilled. Thus, the spot at the waveguide input can be considered as diffraction-limited.\\

\noindent The linear polarization of the CO$_{2}$ laser has been converted into elliptical polarization by adding a quarter-wave plate in the ray path. This permits to force any polarization direction at the waveguide input. The elliptical polarization of the source is characterized by the curve in the left panel of Fig.~\ref{ExpRes} that plots the calibrated flux level of the source as a function of the angular position of the polarizer.\\

\noindent The 300 $\mu$m spacing between the different channel waveguides is used as a qualitative validation criterion for the observation of the waveguide output. Indeed, we can observe successive bright spots when translating the chip by the separation of $300\mu$m, which confirms that the detected flux has been guided along the waveguide.
For quantitative relative flux measurements, a flat field correction is applied and the mean level of thermal background is subtracted as presented hereafter.\\
For each image with a flux $I_{\rm raw}$ per pixel, a dark image $I_{\rm dark}$ corresponding to the response of the array to a spatially uniform source is recorded. The flat field correction is applied for each single pixel by computing a corrected image $I_{\rm flat}$ given by
\begin{eqnarray}
&&I_{\rm flat}=\frac{I_{\rm raw}}{I_{\rm dark}/ \tilde{I}_{\rm dark}}\label{flat_formula}
\end{eqnarray}

\noindent where $\tilde{I}_{\rm dark}$ is the pixel average value of the dark image. The mean level of the thermal background is obtained by computing the pixel average $\tilde{m}_{\rm BG}$ of a sub-frame of $I_{\rm flat}$ where the background is uniform and where no signal is detected. The final corrected image $I_{\rm cor}$ is given by

\begin{eqnarray}
&&I_{\rm cor}=I_{\rm flat}-\tilde{m}_{\rm BG}\label{corrected_formula}
\end{eqnarray}

\noindent As an example, Fig.~\ref{cor} presents a cross-sectional view of the correction stage as well as a calibrated image of a waveguide output. The artifacts visible on the left panel of Fig.~\ref{cor} are filtered out with the applied correction.\\
\noindent Finally, to perform transmission measurements of a waveguide, we compare the calibrated flux without the waveguide $I_{\rm cor}^{\rm out}$ and the calibrated flux through the waveguide $I_{\rm cor}^{\rm in}$.
\noindent The estimated relative transmission of the waveguide is thus
\begin{eqnarray}
&&T_{\rm est}=\frac{I_{\rm cor}^{\rm in}}{I_{\rm cor}^{\rm out}}\label{first}
\end{eqnarray}

\noindent where $I_{\rm cor}^{\rm in}$ is the transmitted power through the waveguide and $I_{\rm cor}^{\rm out}$ is the transmitted power without the waveguide. $T$ is defined as the total throughput of the component.\\
\noindent The time variations of the source and of the the thermal background in the laboratory have been measured over a time-scale of 1 hour to evaluate the achievable signal-to-noise ratio. The characterization has resulted in a variation of 0.2\% for the thermal background and 1.5\% for the laser signal.

\section{Experimental Results}\label{results}

\subsection{Characterization of the modal behavior by polarization tests at $\lambda$=10.6$\mu$m}

\begin{figure}
\begin{minipage}{\columnwidth}
\centering
\includegraphics[width=4.5cm]{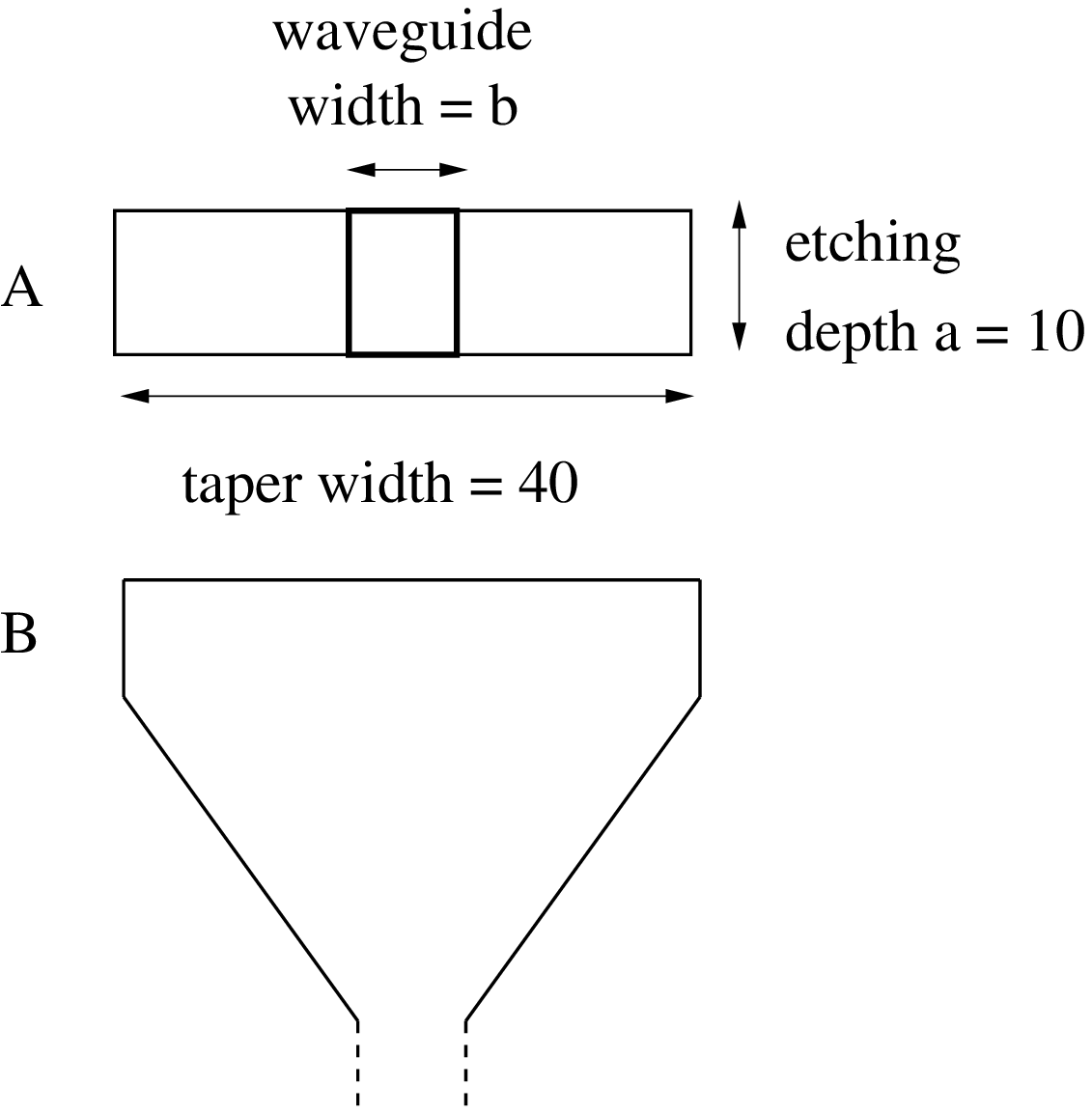}
\hspace{0.1cm}
\includegraphics[width=4.0cm]{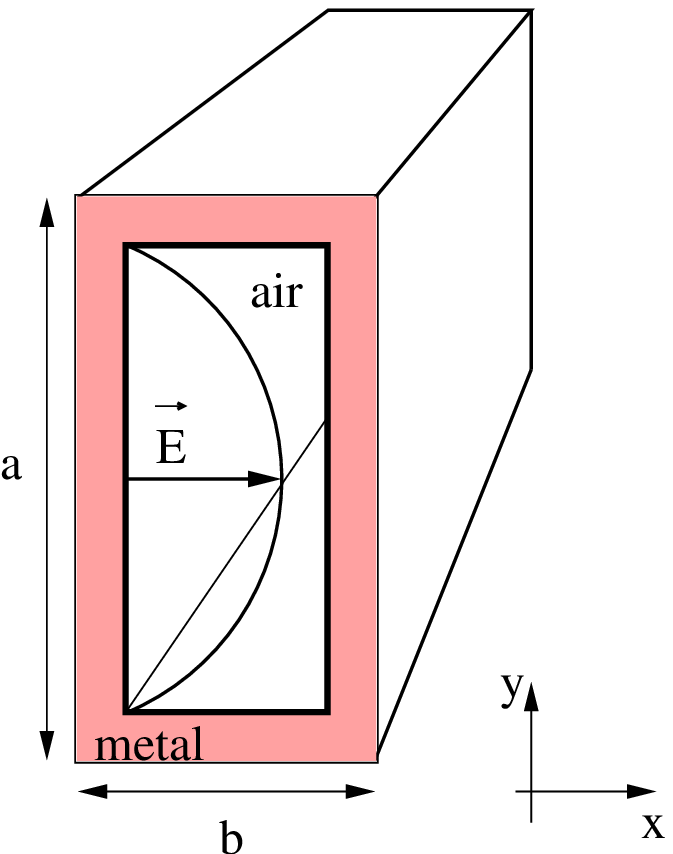}
\end{minipage}
\caption{Left panel: front view (A) and upper view (B) of a taper. The etching depth is fixed to 10 $\mu$m and the taper width equals 40 $\mu$m. The waveguide width varies according to Table~\ref{ztable}. Right panel: geometry of Hollow Metallic Waveguides designed to be single-mode. The dominant mode is the fundamental TE$_{10}$ whose distribution follows a cosine y-dependent distribution law.}\label{Waveguide_Geom}
\end{figure}

As presented in Sect.~\ref{component}, a rectangular HMW designed to be single-mode supports only one state of polarization
while a multi-mode HMW presents at least two orthogonal states of polarization with different field distributions. According to its geometry, the fundamental mode is the TE$_{10}$ with the electric field oriented along x-axis (see Fig.~\ref{Waveguide_Geom}) and the potential first higher order mode is the TE$_{01}$ with the electric field oriented along the y-axis \citep{Jordan}. To test the modal behavior of the waveguides, we have placed the grid polarizer after the component output to analyze straightforward the polarization state. Depending on if a waveguide is single-mode or multi-mode, flux will be detected for any angular position (multi-mode) or only for specific angular position (single-mode) of the polarizer. Analyzing the output polarization state is a simple method that gives unbiased information to discriminate between TE$_{10}$ and TE$_{01}$ propagating modes.\\

\begin{figure*}
\begin{minipage}{\textwidth}
\centering
\includegraphics[width=5.0cm]{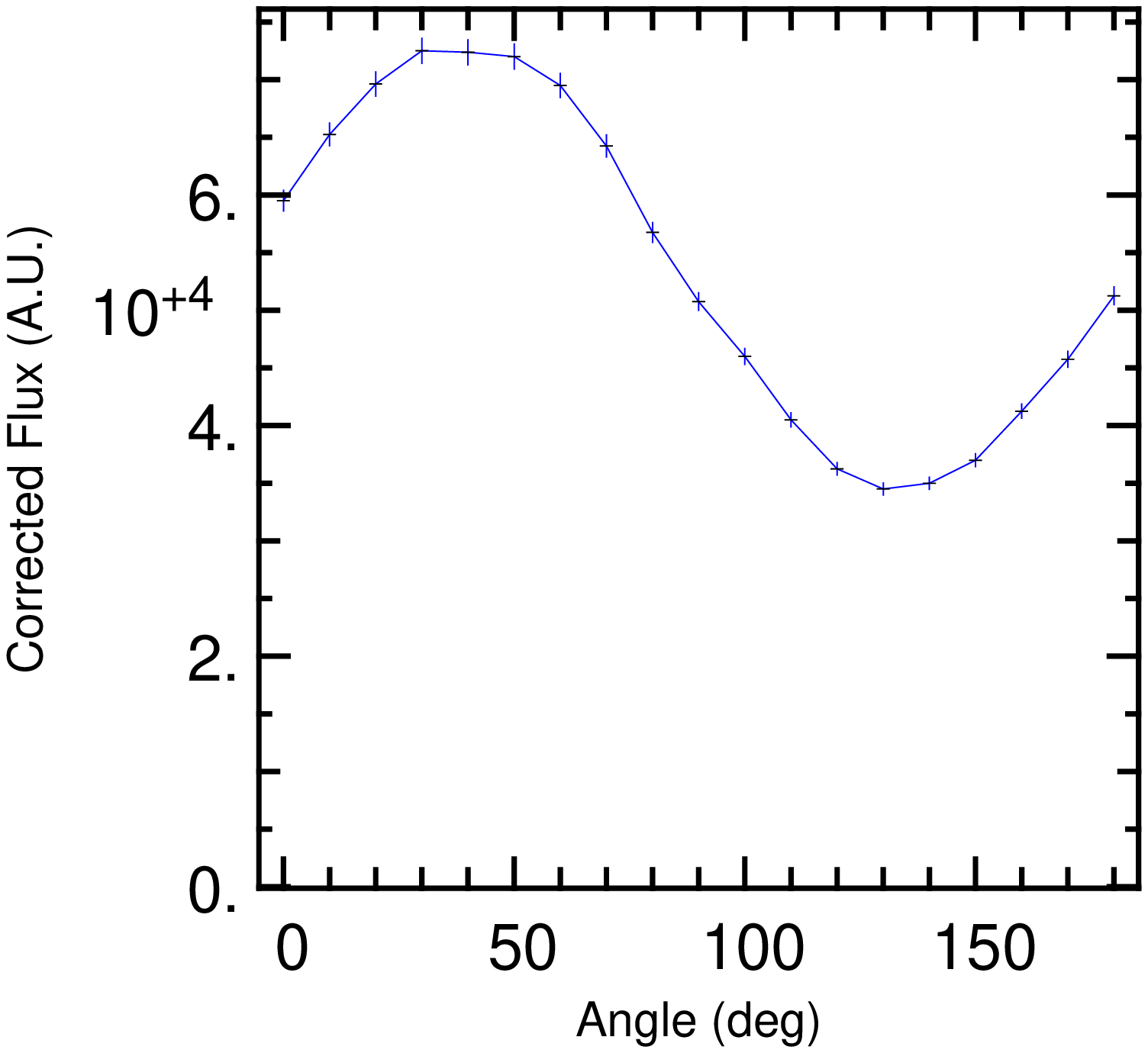}
\hspace{0.5cm}
\includegraphics[width=4.85cm]{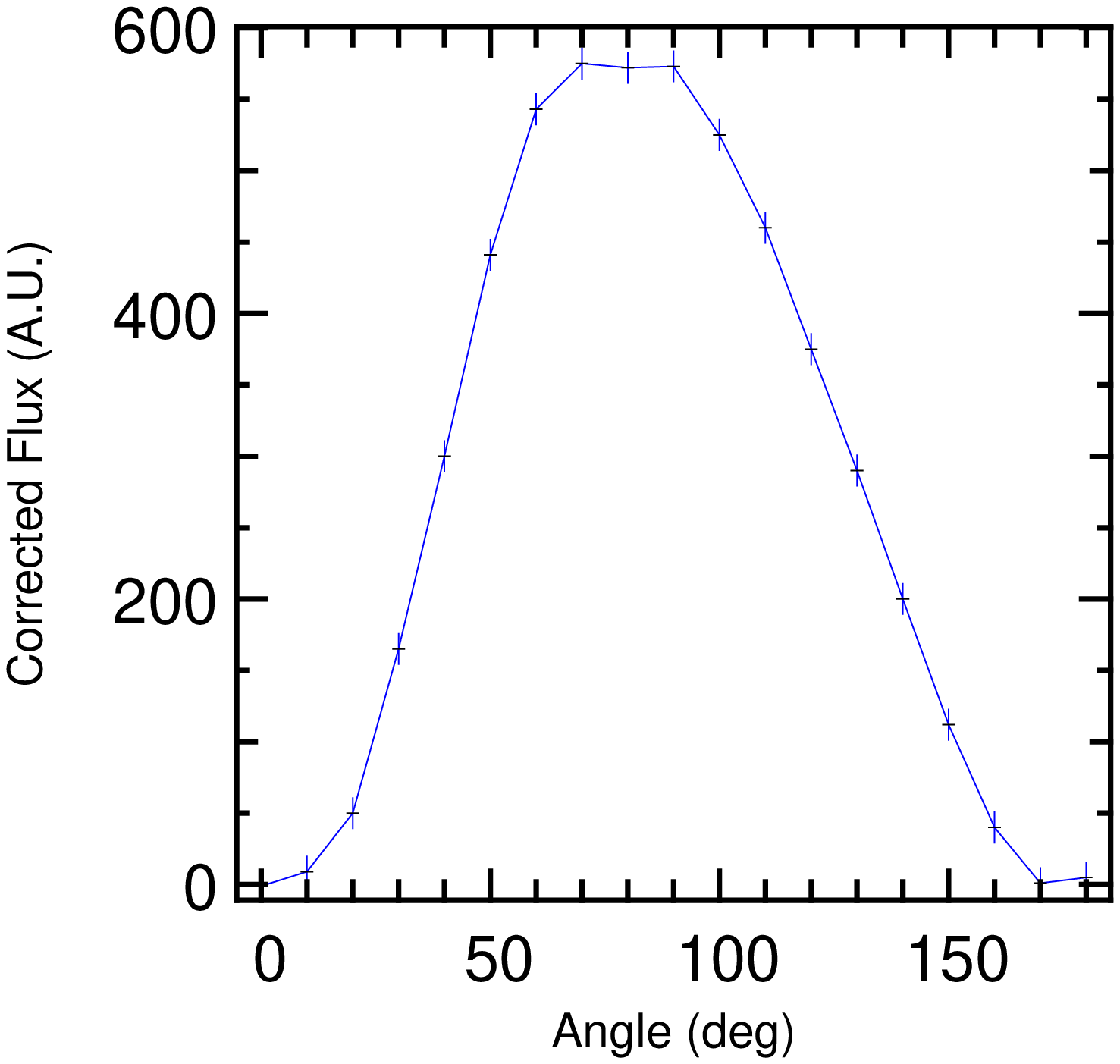}
\hspace{0.5cm}
\includegraphics[width=5.0cm]{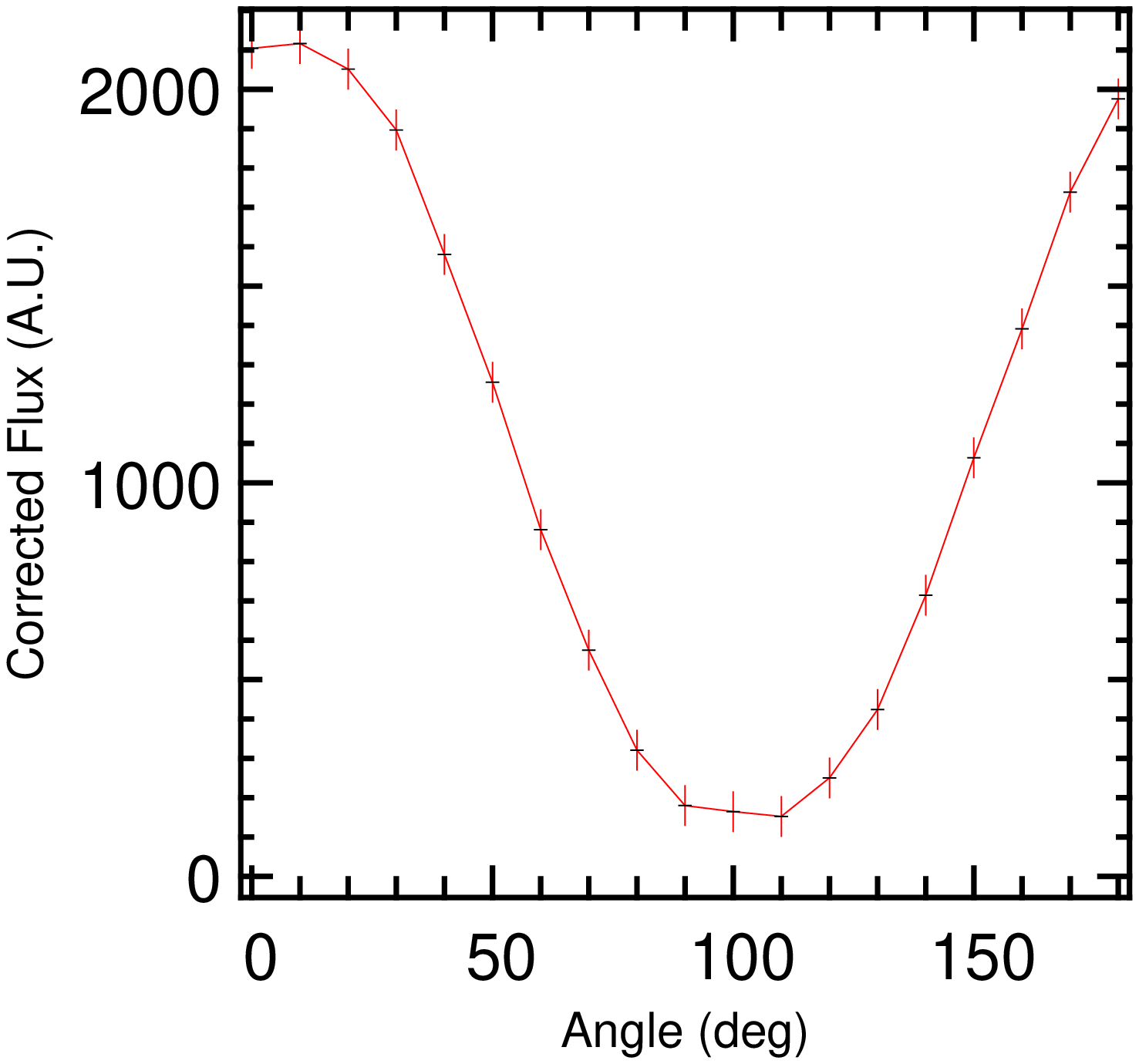}
\end{minipage}
\caption{a: Characterization curve of the elliptically polarized source. - b: Polarization measurements for the single-mode waveguide G36-T at $\lambda$=10.6 $\mu$m with $a$=10 $\mu$m and $b$=5.1 $\mu$m. The output flux can be totally extinguished at angular positions 0$^{\circ}$ and 180$^{\circ}$ with $\sim$2.5\% accuracy - c: Polarization measurements for the multi-mode waveguide G42-T at $\lambda$=10.6 $\mu$m with $a$=10 $\mu$m and $b$=9.2 $\mu$m. A strong variation of the flux level is observed but it is not possible to achieve a total extinction of the output flux for any of the polarization direction. For the three graphs, the flux level is reported in Arbitrary Units (A.U.).}\label{ExpRes}
\end{figure*}

\noindent The theoretical expected intensity distribution at a single-mode waveguide output is \citep{Jordan}
\begin{eqnarray}
&&I_{\rm{out}}(\theta) \propto I_{\rm{0}}\sin^{2}(\theta)\label{polar_formula}
\end{eqnarray}

\noindent where $I_{\rm{0}}$ is a constant intensity and $\theta$ the angle between the y-axis and the tested polarization direction. The plot in the center panel of Fig.~\ref{ExpRes} shows the calibrated flux in arbitrary units measured after the waveguide as a function of the angular position of the grid polarizer. This result is obtained with the sample G36-T (see Table.~\ref{ztable}) designed to be single-mode at $\lambda$=10.6 $\mu$m.

\noindent The angular positions $0^{\circ}$ and $180^{\circ}$ correspond to the polarizer oriented along the y-axis in Fig.~\ref{Waveguide_Geom}, which theoretically matches with the total extinction of the electric field oriented orthogonally. The plot shows that the electric field can be experimentally totally extinguished at these two extreme angular positions. The first higher order mode being the TE$_{01}$ according to the geometry of G36-T, this result shows that no electrical field oriented along the y-axis can be excited at a wavelength of 10.6 $\mu$m, which confirms the single-mode behavior of the waveguide.\\
\\
\noindent Polarization tests have been also made for waveguides designed to be multi-mode at $\lambda$=10.6 $\mu$m. The results on the tested sample G42-T (see Table.~\ref{ztable}) have been reported here. The propagation modes theoretically existing in the cavity are the fundamental mode TE$_{10}$ with its electric field oriented along the x-axis, the mode TE$_{01}$ with the electric field oriented along the y-axis, and the modes TE$_{11}$ and TM$_{11}$ with a distribution of the electric field both in $\it{x}$ and $\it{y}$ directions \citep{Jordan}. Since TE$_{11}$ and TM$_{11}$ are odd modes, it is assumed that no energy will be coupled on those modes if the excitation field is centered on the waveguide aperture. The right panel of Fig.~\ref{ExpRes} shows the flux measured after the sample G42-T as a function of the angular position of the polarizer.\\
The curve presents a strong decrease around 100$^{\circ}$, but for any position of the polarizer between 0$^{\circ}$ and 180$^{\circ}$ the result shows that it is always possible to detect flux transmitted through the waveguide. The minimum value is about 150, which is above the noise limit. Thus, no single-polarization state can be measured with G42-T, which shows that higher order modes can propagate into the waveguide.\\
For the single-mode G36-T component, the error bars measured for the total extinction are limited by the thermal background fluctuations ($\sim$16 A.U. in Fig.~\ref{ExpRes}). The maximum flux level is about 600 A.U., which guarantees an accuracy of $\sim$2.5\% on the measurement.
It is also interesting to note that the maximum flux detected with the multi-mode waveguide G42-T ($\sim$2100 A.U.) is three to four times the maximum flux detected with the single-mode waveguide G36-T ($\sim$600 A.U.). This can be explained by power coupling on higher order modes that can propagate into the multi-mode waveguide while they are filtered by the single-mode waveguide.\\
\noindent In Table~\ref{ztable} are reported the results of the modal characterization of a precise set of waveguides at $\lambda$=10.6 $\mu$m. The modal behavior of the samples is given based on the results of polarization tests. Our results show that a cut-off is experimentally observed for waveguide widths above $\it{w}$=5.1 $\mu$m. This confirms the single-mode range predicted theoretically in Sect.~\ref{component}.\\
\\
\noindent The analysis based on polarization tests offers a quantitative measurement of the modal behavior that is based on flux level measurements. This method can be compared to the near-field imaging method used in \citet{Laurent} to characterize fibers for the near-infrared, which is based on a qualitative analysis of the output intensity distribution. With this last method, the discrimination between the fundamental and the first mode turns to be more difficult when analyzing the field spatial distribution rather than implementing a flux measurement.

\subsection{Waveguide Transmission}

In a simple approach, the total throughput $\it{T}$ of a channel hollow metallic waveguide  can be separated into 1) input coupling efficiency $\it{C}_{\rm{in,i}}$ on the mode {\rm i} 2) propagation losses $\it{P}_{\rm{i}}$ of the mode {\rm i} 3) output coupling efficiency $\it{C}_{\rm{out,i}}$ on the mode {\rm i} 4) quality factor $\eta_{\rm{i}}$ through the expression
\begin{eqnarray}
&&T=\sum_{\rm{i}=1}^{\rm n} C_{\rm{in,i}}.P_{\rm{i}}.C_{\rm{out,i}}.\eta_{\rm{i}}\label{throughput}
\end{eqnarray}
\noindent which simply becomes in the single-mode case
\begin{eqnarray}
&&T(\%)=C_{\rm{in}}.P.C_{\rm{out}}.\eta\label{throughput_2}
\end{eqnarray}

\noindent The term $P$ is dependent on the waveguide length.
The quality factor $\eta$ includes a term of impedance mismatch comparable to Fresnel losses for near-infrared fibers and the losses due to technological imperfections at the taper input and output that degrade the coupling efficiency and that are not considered when computing the overlap integral. $\eta$ is dependent on the fabrication process and can vary from one waveguide to another.
The term $P.\eta$ is defined as the excess losses and corresponds to the physical quantity that can be measured experimentally. It is therefore a worst-case value of the propagation losses $P$ of the waveguide.\\
The input and output coupling efficiencies must then be evaluated to estimate the propagation term $P.\eta$. They have been estimated numerically by computing the two-dimension overlap integral between the excitation field and the dominant mode profiles at the waveguide input and output.
The excitation field at the waveguide input is a plane wave focused by a lens with focal length {\it f} and a clear aperture {\it D}. The shape of the field can be modelized with an Airy function given by

\begin{eqnarray}
&&E_{\rm{exc}}(x,y,\frac{f}{D}) = 2\frac{J_{1}(\frac{D}{f}.\frac{\pi \sqrt{(x-d)^{2}+(y-e)^{2}}}{\lambda})}{(\frac{D}{f}.\frac{\pi \sqrt{(x-d)^{2}+(y-e)^{2}}}{\lambda})}\label{Bessel}
\end{eqnarray}

\noindent where $E_{\rm{exc}}$ stands for the electric excitation field amplitude, {\it f} is the focal length of optics I$_{1}$ (see Fig.~\ref{layout}) and {\it D} its clear aperture. $\it{J}_{1}$ is the first order Bessel function, $\it{d}$ and $\it{e}$ are the lateral displacement with respect to the center of the waveguide aperture respectively in the $\it{x}$ and $\it{y}$ directions. Since the $E_{\rm{exc}}$ has a radial symmetry and the waveguide aperture two symmetry axis, it can be shown that the maximum coupling efficiency occurs when the excitation field is centered in the waveguide input (i.e $d$=$e$=0). Losses $\le$1\% occur for a displacement $d$ or $e$ of 1-$\mu$m  \citep{Schanen}. Thus, in the next sections the coupling efficiencies are considered to be computed for a centered excitation field.\\

\noindent Due to the quality factor $\eta$, the experimental coupling efficiency might differ from the theoretical values. It is possible to evaluate the order of variation between experience and theory by changing the diameter (i.e. the numerical aperture) of the injection spot given in Eq.~\ref{Bessel} and measuring the variations of the transmitted flux.

\begin{figure}
\centering
\includegraphics[width=5.5cm]{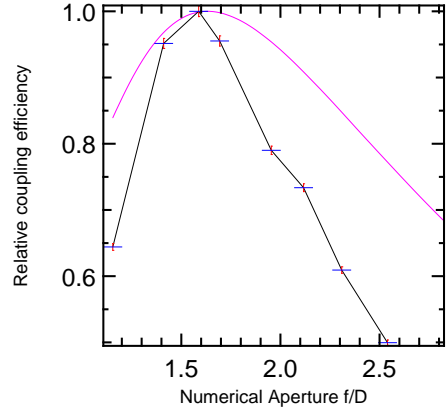}
\caption{Relative coupling efficiency as a function of the incident beam aperture. The cross curve gives the experimental results and the line curve is obtained with the model.}\label{coupling_eff}
\end{figure}

\begin{table*}
\centering
\begin{tabular}{l c c c c c c c}
\hline\hline
 Reference     & b              &    Modal     &  $C_{\rm{in}}$  & $C_{\rm{out}}$ &  $T_{\rm{exp}}$ &  Excess       &  Error (dB)  \\
               & ($\mu\rm{m}$)  &    behavior  &  (\%)           & (\%)           &  (\%)           &  losses (dB)  &              \\ \hline
 $G46-1$       &  4.7           &    SM        &  15.5           & 4.3            &  0.26           &   $<$ -4.0    &  0.13         \\
 $G46-2$       &  4.7           &    SM        &  15.5           & 4.3            &  0.38           &   $<$ -2.4    &  0.14        \\
 $G46-\rm{T}$  &  4.7           &    SM        &  20.9           & 21.3           &  1.9            &   $<$ -3.7    &  0.13        \\ \hline
 $G36-\rm{T}$  &  5.1           &    SM        &  20.9           & 21.3           &  1.8            &   $<$ -3.9    &  0.13        \\ \hline
 $G35-\rm{T}$  &  7.4           &    MM        &   -             &  -             &  4.9            &   -           &  -           \\ \hline
 $G42-\rm{T}$  &  9.2           &    MM        &   -             &  -             &  4.2            &   -           &  -           \\ \hline
 $G63-1$       &  8.4           &    MM        &   -             &  -             &  1.9            &   -           &  -           \\
 $G63-\rm{T}$  &  8.4           &    MM        &   -             &  -             &  5.1            &   -           &  -           \\ \hline
\end{tabular}
\caption{Description of tested waveguides and experimental performances over 1mm length. The sign "-" in the table means the computation is not applicable. The waveguides supporting a taper are referred as Gxx-T. The modal behavior has been assessed through polarization tests. For $b\le$5.1$\mu$m, the transmitted flux can be totally extinguished at 0$^{\circ}$ angular position of the polarizer. Above this width, flux variations are still observed but it is not possible to extinguish the flux anymore. In the single-mode case (SM), the coupling efficiency on the fundamental mode has been computed with the overlap integral method for the different geometrical apertures. In the case of waveguides with tapers (G46-T and G36-T), the coupling efficiencies have been computed for the taper aperture 40$\mu$m$\times$10$\mu$m, which results in similar values of $C_{\rm{in}}$ and $C_{\rm{out}}$. The excess losses include the propagation losses of the waveguide and the effects of the quality factor $\eta$. Using a taper improves significantly the coupling both in single-mode and multi-mode case (MM). Multi-mode waveguides with taper transmit about 4.5\% of the total flux while single-mode waveguides with tapers transmit about 2\% of the total flux.}\label{ztable}
\end{table*}

\noindent The plot of Fig.~\ref{coupling_eff} compares the normalized coupling efficiency on an aperture with a taper obtained with the model approach and with the experimental data. The coupling efficiency curve has been normalized to one because it is only possible to make a relative comparison of the different data via differential flux measurements. It is not possible to extract the absolute values of the efficiencies with our current test configuration.
Doing this test with a tapered waveguide has the advantage of observing experimentally a maximum for the coupling efficiency curve, which would have not be possible testing a waveguide without taper for which a maximum occurs below $f$/0.5.
\noindent The cross-curve first shows that a maximum coupling efficiency is obtained experimentally for $f/D$$\approx$1.6, which matches with the value computed with the model. This results tells that the method of the overlap integral can be applied in our case. Aside from the maximum coupling, the experimental curve presents a significant degradation - up to 20\%  - with the theory. That puts on evidence the important effect of the quality factor $\eta$ in degradation of the coupling efficiencies.\\
\\
\noindent We report in Table~\ref{ztable} the results of the characterization phase of HMW at $\lambda$=10.6$\mu$m.
In Eq.~\ref{throughput_2}, $T$ is measured experimentally, $C_{\rm{in}}$ and $C_{\rm{out}}$ are obtained numerically which permits to extract the excess losses $P.\eta$. The excess losses is the only transmission feature that can be extracted from experimental measurements and is therefore given in dB while the propagation losses are given in dB.mm$^{-1}$.\\
The output coupling efficiency has been computed with the following parameters: the imaging optical system has a focal length $f$=38.1mm and a clear diameter $D$=16.5mm, which gives a numerical aperture $f$/2.29. At $\lambda$=10.6$\mu$m, the imaging optics is limited to 28$\mu$m resolution and a typical aperture of 10$\mu$m$\times$10$\mu$m will not be resolved. This results in the diffraction-limited energy distribution shown in Fig.~\ref{cor}. As a consequence, a Bessel distribution is assumed with the numerical aperture of $f$/2.29 to compute the output coupling efficiency $C_{\rm{out}}$.\\
The injection tests have been carried out with the workbench configuration presented in Sect.~\ref{setup} in which the injection lens I$_{1}$ has a numerical aperture of $f$/1.15. Although this aperture is not the optimal one when considering 40$\mu$m$\times$10$\mu$m tapers, it is the best option for waveguides without tapers.

\noindent For multi-mode waveguides, it is not possible to separate experimentally the excess losses of each mode with our setup. However it is still possible to compare the total throughput as it is done in Table~\ref{ztable}. When considering tapered waveguides, a clear transition can be observed in terms of transmission as predicted by the theory. Filtering the higher order modes results in significant drop in flux, which corresponds to the transition between single-mode and multi-mode regime. The same aspect is encountered with traditional dielectric infrared fibers.\\
\\
\noindent Theoretically, propagation losses for rectangular
single-mode waveguides with $a$=10$\mu$m are $\sim$ 0.8 dB.mm$^{-1}$
at 10.6$\mu m$ (see \citep{Schanen} and appendix~\ref{A-appendix}). A static conductivity of $\sigma_{\rm{0}}$=4.1$\times$10$^{7}$S.m$^{-1}$ has been assumed for gold coating. Experimentally, excess losses of 2.5 dB to 4 dB over 1-mm are measured in this study. Firstly, this difference results from a poor quality factor at the waveguide input and output that strongly affects the coupling efficiency. The important dispersion in losses for the sample G46 in Table~\ref{ztable} could be explained with a similar dispersion in the technological repeatability, which could be improved in a next technological run. However, the clear distinction in transmission between single-mode and multi-mode waveguides show that the quality factor does not prevent from assessing the modal properties of those waveguides. New considerations on waveguide losses are presented in Sect.~\ref{disc}.

\section{Discussion}\label{disc}

This section discusses the results obtained in this work and proposes improvement perspectives. The following properties of HMW are also detailed: operational spectral range, wavelength-dependent losses and polarization properties.\\
\\
\noindent The analysis through polarization tests allowed showing the single-mode behavior of waveguides at 10.6 $\mu$m and estimating it with $\sim$2.5\% accuracy due to background fluctuations.  This first reported measurement confirms that conductive waveguides are suitable for manufacturing single-mode waveguides in the mid-infrared.\\
\noindent The optical quality of the waveguide input and output facets (included within the factor $\eta$ in the paper) greatly affects the excess losses. At the moment, the chips containing the waveguides are cut out from the wafer using dicing and cleaving processes, which results in unexpected irregularities around the waveguide inputs. Different technological options have already been considered, which should reduce the dispersion on quality and then improve the technological repeatability.\\
\noindent The experimental results have shown the fundamental importance of coupling optimization at the waveguide input and output. In this first technological run, a strong priority has been set on manufacturing rather than on design, so the computed coupling efficiencies in Tab.~\ref{ztable} are not optimized. Thus, we are confident that coupling efficiencies can be improved through a more stringent design of the tapers in the next run.\\
\\
\noindent One of the most recurring issue for spatial or modal filtering is the effects of chromatism.
With optical and infrared waveguides, modal filtering can be
implemented in a spectral range where the device is single-mode. We
have seen that for conductive rectangular waveguides with dimensions
$a$ and $b$=$a$/2, the single-mode domain is given for $a$ $\le$
$\lambda$ $\le$ 2$a$, which represents a 66\% bandwidth centered on
$\lambda$ = 1.5a $\mu$m. Thus, this property is used to divide
  the [4 - 20 $\mu$m] band into a relatively low number (three or
  four) of single-mode sub-bands.\\
However, it is not recommended to apply this reasoning straightforward
since the propagation losses within the same sub-band are highly
chromatic. \citet{Schanen} has derived the well-established losses
calculations in millimeter regime to the mid-infrared domain using a
Drude's modal approach for the metallic
coating. Appendix~\ref{A-appendix} presents the main steps of the
calculation for this approach. The author has expressed the attenuation coefficiant of the dominant mode TE$_{\rm 10}$ as a function of $\lambda$, $\lambda_{\rm c}$, and ($n$,$k$) the real and imaginary indeces of gold in the mid-infrared. The conclusion contains the following two points: first, the average losses in a given single-mode sub-band are higher at shorter wavelength than at longer ones; secondly, within the same sub-band, losses are minimized for wavelengths close to the lower cut-off (i.e. $\lambda_{\rm{c}}$ = $a$) and increase continuously when $\lambda$ approaches the higher cut-off (i.e. $\lambda_{\rm{c}}$ = 2$a$) \footnote{The attenuation coefficient of a propagating mode tends theoretically to infinity when $\lambda$ equals the cut-off wavelength.}. Dividing the range [4 - 20 $\mu$m] is therefore a trade-off between the bandwidth and the average losses of a single-mode sub-band.\\

\noindent Following this analysis, a trade-off is proposed dividing the [4 - 20 $\mu$m] band into four single-mode sub-bands B$_{\rm 1}$=[4 -- 6 $\mu$m], B$_{\rm 2}$=[6 -- 10 $\mu$m], B$_{\rm 3}$=[10 -- 16 $\mu$m] and B$_{\rm 4}$=[16 -- 20 $\mu$m] using as modal filters HMW with $a$  verifying $a$=4 $\mu$m for B$_{\rm 1}$, $a$=6 $\mu$m for B$_{\rm 2}$, $a$=10 $\mu$m for B$_{\rm 3}$ and $a$=16 $\mu$m for B$_{\rm 4}$\footnote{The fourth waveguide is single-mode up to $\lambda$ = 32 $\mu$m, but the definition of the Darwin band limits it to $\lambda$ = 20 $\mu$m.}. The computed average losses for each sub-bands are respectively 3.5 dB.mm$^{-1}$ for B$_{1}$, 2.7 dB.mm$^{-1}$ for B$_{2}$, 1.5 dB.mm$^{-1}$ for B$_{3}$ and 0.6 dB.mm$^{-1}$ for B$_{4}$.\\

\noindent Compared to the computed performances, the transmission values presented in Table~\ref{ztable} are also affected by the surface roughness of the metallic walls, which is not taken into account yet in the loss model. To date a gold coating with rugosity lower than 50nm rms can be deposited but the roughness has increased during the anodic bonding process. A better control of this last step is achievable, making roughness lower.\\
\citet{Schanen} has shown that an efficient filtering of 10$^{-7}$
($\sim$ 70 dB) is already reached after a 10$\lambda$ length ($\sim$ 100
$\mu$m) of the waveguide (see appendix~\ref{B-appendix}). This implies that integrated optics
functions (e.g. IO beam combiner) could be implemented over short
propagation distances, typically lower than 1-mm length, reducing then
the propagation losses. Fabricating shorter functions is a point that
has already been addressed and is technically achievable.\\
\\
Finally, HMW present interesting polarization properties with respect
to the issue of polarization control presented in
Sect.~\ref{intro}. A single-mode rectangular metallic waveguide
presents a single-polarization behavior, which means the
implemented device is able to force a linear polarization at its
input. Under those conditions, the beams combined in the waveguide do
not present polarization mismatching anymore. Since the differential
rotation of polarization planes has to be controlled down to 0.002 rad
to achieve a 10$^{6}$ rejection ratio \citep{Mennesson},
single-polarization waveguides present an advantage to get rid of this
constraint although this could occur at the expense of power
transmission.


\section{Conclusions}

This study has focused on the feasibility of manufacturing rectangular
Hollow Metallic Waveguides to be used as modal filters thanks to their
single-mode behavior. This offers an original instrumental solution
in a spectral range where similar possibilities are only recently
emerging. The demonstrated single-mode behavior is considered as the
major result of this study.\\
A quantitative measurement based on experimental polarization analysis has shown the single-mode behavior of the manufactured waveguides with $\sim$2.5\% accuracy and has been compared with results obtained on multi-mode waveguides. Single-mode HMW show that a single-polarization state can be maintained into the waveguide, which is a certain advantage for polarization control in nulling interferometry.
The result of this study also states that improving the total throughput of such waveguides is mainly a matter of increasing the coupling efficiencies at waveguide inputs and outputs. This can be increased significantly through the designing of well-adapted horns.\\
In a medium-term scale, single-mode channel waveguides could be extended to the manufacturing of planar integrated components. Those devices could ensure simple interferometric functions (recombination, photometric control), which compactness and stability would be very attractive for future interferometry space missions.

\appendix
\section{Deriving of the theoretical propagation losses of the
  TE$_{10}$ mode}\label{A-appendix}

This annex presents the main results contained in \citet{Schanen}.
In the microwave regime -- i.e. for frequencies ranging from 1 to 1000
GHz -- the computation of the theoretical propagation losses is based
on the modelization of the skin ``skin effect'' \citep{Rizzi}. It is
well-known that the tangential electric field at the surface of a
perfect metallic medium equals zero. For a real metallic coating, the
finite conductivity induces the presence of a small tangential
component of the electric field in the metal. As a consequence, some
power flows through the lateral walls of the waveguide.\\
To extract the attenuation factor of the guided mode, we identify the
average guided power lost in between the coordinates $z$=$0$ and $z$=$L$ of
the propagation axis with the power flowing through the metallic
walls on an equivalent length. In this approach, the static conductivity of the metal
$\sigma_0$ has to be considered as long as we remain in the microwave
regime since the frequency of the electromagnetic wave is
much lower than the collision frequency of the electrons ($\sim$
10$^{13}$ Hz \citep{Quere}). In the mid-infrared domain domain, the
radiation frequency is much higher ($\sim$ 3$\times$10$^{13}$ Hz) and
the conductivity of the metal $\sigma$ becomes
frequency-dependent following the Drude model \citep{Quere}. This has
been taken into account in the proposed approach.\\
The power dissipated in the walls $W_J$ between $z$=$0$ and $z$=$L$ is
obtained by computing the flux of the Poynting vector across the
waveguide section at the corresponding $z$-coordiantes. This results
in

\begin{eqnarray}
W_{J} &=& (1-\exp(-2\alpha_{\rm g} L))\int_{section}(\vec{E} \times
\vec{H}^{*}).\vec{z}dxdy\label{A-lostpower}
\end{eqnarray}

\noindent where $\vec{E}$ and $\vec{H}$ are respectively the electric and
magnetic field, $\alpha_{\rm g}$ the attenuation factor. Using the equations
of $\vec{E}$ and $\vec{H}$ fields for the TE$_{10}$ fundamental mode
\citep{Jordan}, Eq.~\ref{A-lostpower} becomes 

\begin{eqnarray}
W_{J} &=& (1-\exp(-2\alpha_{\rm g}
L))C^{2}\frac{2}{\lambda^{2}}ba^{3}\sqrt{\frac{\mu_0}{\epsilon_0}}\sqrt{1-\frac{\lambda^{2}}{\lambda_{c}^{2}}}
\label{A-lostpower2}
\end{eqnarray}

\noindent where $C$ is a field constant, $\lambda$ the operating
wavelength, $\lambda_{c}$ the TE$_{10}$ cut-off wavelength, $a$ and
$b$ the waveguide dimensions.\\
To compute the tangential component of the $\vec{E}$ field in the
metal, we employ the Maxwell equation 

\begin{eqnarray}
\frac{d}{dz}H_{x}-\frac{d}{dx}H_{z} & = & -\jmath\omega\epsilon
E_{y}+\sigma E_{y}\label{A-maxwell1}
\end{eqnarray}

\noindent where $\epsilon$ and $\sigma$ are the complex permittivity
and conductivity of the metal obtained through the Drude model of a
metal, $omega$ the radiation frequency. At infrared wavelengthes, the commonly used approximation
Re[$\sigma$] $>>$ Re[$\epsilon$]$\omega$ is not valid anymore and both terms must be maintained in
Eq.~\ref{A-maxwell1}.
The wave vector $\vec{k}$ of TE$_{10}$ mode presents a
component $k_z$ along the propagation axis, as expected in the case
of a perfect metal, and a component along the $y$-axis (see
Fig.~\ref{Waveguide_Geom}) corresponding to a slight penetration of
the $\vec{E}$ field in the metal. Quantities $k_z$ and $k_x$ are
determined using the formalism of the reflection of an electromagnetic
radiation at a metallic surface. We derive

\begin{eqnarray}
k_{z}&=&\jmath\frac{\omega}{c}\sqrt{1-\left(\frac{\lambda}{\lambda_c}\right)^2}\\
k_{x}&=&\jmath\frac{\omega}{c}\sqrt{(n-\jmath\kappa)^2-(1-\left(\frac{\lambda}{\lambda_c}\right)^2)}
\end{eqnarray}

\noindent $N$=$n-\jmath\kappa$ is the complex refractive index of the
metal. Quantities $n$ and $k$ are accessible at $\lambda$=10 $\mu$m
through measurement tables in \citet{Ordal83}. Solving
Eq.~\ref{A-maxwell1}, we obtain an expression of the $\vec{E}$ field
in the metal following 

\begin{eqnarray}
E_{x} & = & C\frac{1}{\jmath\omega\epsilon-\sigma}\frac{k_{y}^{2}+k_{z}^{2}}{k_y}\exp(-\jmath\omega t)\exp(k_{y}y+k_{z}z)
\end{eqnarray}

\noindent The expressions of the conductivity $\sigma$ and permittivity
$\epsilon$ follow

\begin{eqnarray}
\sigma & = & \frac{\sigma_0}{(1-\jmath\omega\tau)}\label{A-drude1}\\
\sigma_0 & = & \epsilon_0\epsilon_r\tau(\omega_p)^2\label{A-drude2}
\end{eqnarray}

\noindent where $\omega_p$ is called ``plasma frequency'' and
caracterizes the media \citep{Ordal83}. The power flowing through the
metallic walls is obtain in an analog way of Eq.~\ref{A-lostpower}
through

\begin{eqnarray}
P_{J}&=&\frac{1}{2}Re[\sigma]\int_{V}\vec{E_x}.\vec{E_x}^{*}dV\label{A-powerlost2}
\end{eqnarray}

\noindent where $Re$[ ] is the ``real part'' operator. The quantity $dV$ is the elementary volume that takes into
account the walls surface and the penetration depth of the $\vec{E}$
field in the metal. We derive the following expression for $P_J$

\begin{eqnarray}
\lefteqn{
P_{J}{}={}
\frac{1}{2}C^{2}\frac{Re[\sigma]}{\sigma^{2}+(\omega\epsilon)^2}{}\left|\frac{(k_x)^2+(k_z)^2}{k_x}\right|^{2}{}}
\nonumber \\ & &
{}\times{}{}b.\frac{1}{2Re[k_x]}\frac{1-\exp(-2\alpha_{\rm
    g}L)}{2\alpha_{\rm g}}\label{A-powerlost3}
\end{eqnarray}

\noindent From Eq.~\ref{A-drude1} and~\ref{A-drude2}, the term
$(Re[\sigma]/(|\jmath\omega\epsilon-\sigma|^2))$ can be simplified
into $1/\sigma_0$, where $\sigma_0$ is the static conductivity of the
metal. Using the fact that 

\begin{eqnarray}
\kappa^2-n^2 > 1 > 1-\left(\frac{\lambda}{\lambda_c}\right)^2
\end{eqnarray}

\noindent we derive the following expressions 

\begin{eqnarray}
Re[k_x] & = &
\frac{\omega_c}{c}\kappa\left(1+\frac{\left(1-\left(\frac{\lambda}{\lambda_c}\right)^2\right)}{2(n^2+\kappa^2)}\right)\label{A-final1}
\end{eqnarray}

\noindent and

\begin{eqnarray}
\lefteqn{
\left|\frac{(k_x)^2+(k_z)^2}{k_x}\right|^2{}={}} \nonumber \\ & &
{}\frac{\omega_c}{c}\left(n^2+\kappa^2+2\left(1-\left(\frac{\lambda}{\lambda_c}\right)^2\right)\frac{(n^2-\kappa^2)}{(n^2+\kappa^2)}\right)\label{A-final2}
\end{eqnarray}

\noindent Eq.~\ref{A-final1} and~\ref{A-final2} are used to derive the
complete analytical expression of Eq.~\ref{A-powerlost3}. The
attenuation factor $\alpha_{\rm g}$ is now obtained by assuming that
$W_J$=2$P_J$. The attenuation factor, given in Np.m$^{-1}$ (Nepers per
meter) verifies

\begin{eqnarray}
\lefteqn{
\alpha_{g} {}={}
\frac{\pi}{2}\frac{1}{\sigma_0}\sqrt{\frac{\epsilon_{0}}{\mu_{0}}}\left(\frac{n^2+k^2+2\left(1-\left(\frac{\lambda}{\lambda_c}\right)^2\right)\frac{n^2-k^2}{n^2+k^2}}{k\left(1+\frac{1-\left(\frac{\lambda}{\lambda_c}\right)^2}{2\left(n^2+k^2\right)}\right)}\right){}{}}
\nonumber \\ & & 
{}\frac{\lambda}{a^3}\frac{1}{\sqrt{1-\left(\frac{\lambda}{\lambda_c}\right)^2}}\label{A-zfinal}
\end{eqnarray}

\noindent Eq.~\ref{A-zfinal} is converted into more useful units
dB.m$^{-1}$ by multiplying the expression by factor 8.68 \citep{Jordan}

\section{Deriving of filtering length for higher order modes
  suppression}\label{B-appendix}

\noindent We consider a rectangular HMW with $a$=2$b$. The single mode
domain ranges from $\lambda$=$a$ to $\lambda$=2$a$. The cut-off
wavelenght of the first higher order mode TE$_{01}$ is
$\lambda_c$=$a$=2$b$.\\
\noindent In the single-mode regime, the propagation constant $\gamma$
of the TE$_{01}$ mode as it appears in $\exp({-\gamma z})$ is real and
equals

\begin{eqnarray}
\gamma & = & \frac{2\pi}{\lambda}\sqrt{\left(\frac{\lambda}{\lambda_c}\right)^{2}-1}\label{B-annex1}
\end{eqnarray}

\noindent with $\lambda$$>$$\lambda_c$. The term $\gamma$ is transformed
into attenuation factor $\alpha$ of the mode in dB.m$^{-1}$ through
\citep{Jordan}

\begin{eqnarray}
\alpha & = & 8.68 \times \gamma\label{B-annex2}
\end{eqnarray}

\noindent Assuming a HMW with dimensions $a$=2$b$=10$\mu$m, the
attenuation $\alpha$ factor is lower as $\lambda$ is closer to
$\lambda_c$. Let us consider a worst case with $\lambda$=10.1 $\mu$m.
At this wavelength, Eq.~\ref{B-annex1} provides an attenuation factor
that equals $\alpha$=0.76 dB.$\mu$m$^{-1}$. Thus the waveguide length
required to attenuate the higher order mode by 70 dB ($\sim$10$^{-7}$)
is 92 $\mu$m. As the wavelength increases, the attenuation factor
increases as well and the required filtering length becomes shorter.

\begin{acknowledgements}
      This work was supported by a \emph{European Space Agency} ($\it
      {ESA}$) under contract $16847/02/NL/SFe$ and supported by
      funding from the \emph{French Space National Agency} ($\it
      {CNES}$) and {\it Alcatel Space}.
\end{acknowledgements}

\bibliographystyle{aa}
\bibliography{./bibtex/paper_gc.bib}

\end{document}